\documentclass{article}  
\usepackage{geometry}
\usepackage[nocompress]{cite}

\title{\LARGE \bf
Predictable Interval MDPs through Entropy Regularization
}

\author{M.J.T.C.\ van Zutphen, G.\ Delimpaltadakis, W.P.M.H.\ Heemels, and D.J.\ Antunes
\thanks{This research is part of the research program SYNERGIA (project number 17626), which is partly financed by the Dutch Research Council (NWO).}
\thanks{The authors are with the Control Systems Technology Group, Department of Mechanical Engineering, Eindhoven University of Technology, The Netherlands. \hbox{E-mail:} $\{$\texttt{m.j.t.c.v.zutphen,
i.delimpaltadakis,
m.heemels,
d.antunes}$\}$\texttt{@tue.nl}.}%
}

\usepackage{mathtools}
\usepackage{amsmath,amsfonts,amsthm}
\newtheorem{theorem}{Theorem}
\newtheorem{lemma}{Lemma}
\newtheorem{definition}{Definition}
\newtheorem{remark}{Remark}

\usepackage{algorithm}
\usepackage{algpseudocode}

\usepackage{graphicx}
\usepackage{float}
\usepackage{color}
\usepackage{import}

\begin{document}
\maketitle
\date{}
\thispagestyle{empty}
\pagestyle{empty}

\begin{abstract}
Regularization of control policies using entropy can be instrumental in adjusting predictability levels of real-world systems. Applications benefiting from such approaches range from cybersecurity, which aims at maximal unpredictability, to human-robot interaction, where predictable behavior is highly desirable. In this paper, we consider entropy regularization for interval Markov decision processes (IMDPs), which are uncertain MDPs where transition probabilities are only known to belong to specified intervals. Lately, IMDPs have gained significant popularity in the context of abstracting stochastic systems for control design. In this work, we address robust minimization of the linear combination of entropy and a standard cumulative cost in IMDPs, thereby establishing a trade-off between optimality and predictability. We show that optimal deterministic policies exist, and devise a value-iteration algorithm to compute them. The algorithm solves a number of convex programs at each step. Finally, through an illustrative example we show the benefits of penalizing entropy in IMDPs.

\end{abstract}

\section{INTRODUCTION} \label{sec:introduction}
Since its introduction by Shannon, the concept of information entropy has always been strongly related to Markovian processes  \cite{Cover1991ElementsOI}. Apart from a theoretical interest, entropy optimization is valuable in many practical applications. In stochastic systems, entropy encapsulates the predictability of their behavior, and thus penalizing it makes the resulting system more predictable. In applications such as cybersecurity \cite{Biondi2013MaximizingEO} and surveillance \cite{Guo2023EMPatrollerEM}, larger values of entropy are favorable, as this implies the difficulty for an adversary to predict the behavior of the system. In contrast, for autonomous systems that need to cooperate, be it with humans or other systems, predictability is generally a highly desirable trait~\cite{Ornia2023PredictableRL}.

Minimization of entropy (rate), alongside a reward has recently been investigated in the context of reinforcement learning [\citen{Ornia2023PredictableRL},~\citen{Eysenbach2021RobustPC}], while maximization of policy entropy in reinforcement learning \cite{Haarnoja2018SoftAO} has already had enormous success in practice \cite{Haarnoja2018SoftAA}. Additionally, methods have recently been developed that maximize the entropy and entropy rate of interval Markov chains (IMCs; the generalization of Markov chains (MCs) to interval-valued transition probabilities) [\citen{Biondi2013MaximizingEO},~\citen{Chen2014OnTC}]. Similar research has been conducted for maximization of infinite-horizon entropy [\citen{Savas2018EntropyMF_TL},~\citen{Chen2022EntropyRM}] and its trade-off to cost optimality in Markov decision processes (MDPs) \cite{Bertsekas1995DynamicPA} through expected reward constraints \cite{Savas2018EntropyMF_OG}. 

Despite this recent work, optimization of entropy on interval Markov decision processes (IMDPs) \cite{Givan2000BoundedparameterMD} has not been addressed. IMDPs are uncertain MDPs, where the transition probabilities are only known to belong to action-dependent intervals. IMDPs have recently been receiving considerable attention in many applications [\citen{Hahn2019IntervalMD},~\citen{Mathiesen2024IntervalMDPjlAV},~\citen{Jafarpour2023ACD},~\citen{vanZutphen2023OptimalSP}], especially as abstractions of stochastic systems for formal verification and control design [\citen{Soudjani2013AggregationAC},~\citen{Delimpaltadakis2022IntervalMD},~\citen{Delimpaltadakis2022FormalAO},~\citen{Lahijanian2015FormalVA}]. The IMDP setting raises unique challenges w.r.t.\ IMCs [\citen{Biondi2013MaximizingEO},~\citen{Chen2014OnTC}] and MDPs [\citen{Savas2018EntropyMF_TL},~\citen{Chen2022EntropyRM}]. In fact, due to the action-dependent uncertainty on the transition probabilities in IMDPs, two agents are involved in the robust minimization problem; an agent who aims to minimize the objective function and an adversary that resolves uncertainty in an adversarial manner, maximizing the objective function.

In this work, we address robust minimization of the linear combination of entropy and a standard cumulative cost in IMDPs, thereby establishing a trade-off between optimality and predictability. We formally show the existence of optimal deterministic policies. Note that this property of entropy minimization is not surprising, as the aim is predictability. Further, we devise a value-iteration algorithm that computes a robustly optimal policy (i.e., a policy optimal w.r.t.\ the worst-case adversary) and the corresponding tight upper bound on the linear combination of cumulative cost and entropy. The algorithm solves $|S|\times |A|$ convex programs at each time step, where $|S|$ and $|A|$ are the number of states and actions of the IMDP, respectively. As such, through our algorithm, computation of the optimal policy and the associated upper bound on the combined objective is carried out efficiently, employing convex optimization.

The remainder of the paper is organized as follows. Section \ref{sec:problemformulation} provides the problem formulation. Sections \ref{sec:mainresultI} and \ref{sec:mainresultII} discusses our main results: the value-iteration algorithm in Section \ref{sec:mainresultI}, and the determinism of entropy minimizing policies in Section \ref{sec:mainresultII}. Section \ref{sec:application} provides  a numerical example and Section \ref{sec:conclusion} provides concluding remarks. The proofs of the results are given in Section \ref{sec:proofs}.

\section{PROBLEM FORMULATION} \label{sec:problemformulation}
We start by formally introducing IMDPs in Section \ref{sec:interval_markov_decision_processes}. We then discuss the relationship between Markov processes and Shannon Entropy in Section \ref{sec:entropy_of_markov_processes}. In Section \ref{sec:problem_statement}, we define the problem of finding the cost-entropy upper-bound minimizing policy for IMDPs.

\subsection{Preliminaries: IMDPs, Policies and Adversaries}\label{sec:interval_markov_decision_processes}

Let the set of all discrete probability distributions of size $n\in\mathbb{N}$ be denoted by $\mathsf{P}^n:=\{p\in[0,1]^n \ : \ \sum_{i=1}^{n}p_i=1\}$.
\begin{definition}[IMDP]
    An \emph{interval Markov decision process} (IMDP) is a tuple $
    \mathcal{I}=(S,A,\alpha,c,c_h,\underline{P},\overline{P},h)$,
    with finite state and action spaces, $S$ and $A$, respectively, initial state distribution $\alpha\in\mathsf{P}^{|S|}$, stage cost $c:S\times A\rightarrow \mathbb{R}$, terminal cost $c_h: S\rightarrow\mathbb{R}$, transition probability bounds $\underline{P}: S\times A\times S\rightarrow [0,1]$, $\overline{P}: S\times A\times S\rightarrow [0,1]$, and finite horizon $h\in\mathbb{N}$.
\end{definition}

For all $s,q\in S$ and $a\in A$, it holds that $\underline{P}(s,a,q)\le \overline{P}(s,a,q)$ and $\sum_{q\in S}\underline{P}(s,a,q)\le 1\le \sum_{q\in S}\overline{P}(s,a,q)$. Given a state $s\in S$ and an action $a\in A$, a transition probability distribution $p\in\mathsf{P}^{|S|}$ is called \emph{feasible}, if $\underline{P}(s,a,q)\le p_q\le \overline{P}(s,a,q)$, for all $q\in S$. Let the (convex) set of all feasible distributions for the state-action pair $(s,a)$ be defined as
\begin{equation}
\mathcal{P}_{s}^{a} :=\{p\in\mathsf{P}^{|S|}:\forall q\in S, \underline{P}(s,a,q)\le p_{q}\le \overline{P}(s,a,q)\}.\label{eq:IMDP_p_set}
\end{equation}
While we assume the stage cost and transition probability bounds to be time-invariant, all methods below can straightforwardly be modified to accommodate for time-varying stage cost and transition probability bounds.

\begin{definition}[Policy]
    For an IMDP $\mathcal{I}=(S,A,\alpha,c,c_h,\underline{P},$ $\overline{P},h)$, a policy is defined as a map $\pi : \{0,1,\dots,h-1\}\times S\rightarrow \mathsf{P}^{|A|}$. Hence, a policy $\mu$ is a function that, given the state $s\in S$, at time step $k\in\{0,1,\dots,h-1\}$, produces a probability distribution governing the selection of actions $a\in A$. The set of all policies is denoted by $\Pi$.
\end{definition}

Note that we focus on Markov policies, i.e., the policies that we consider depend only on the present state and not the history of the process. Extensions to non-Markovian policies are left for future work. Nonetheless, it is worth noting that, in most scenarios, Markov policies are indeed sufficient for optimality \cite{Delimpaltadakis2022IntervalMD}.

\begin{definition}[Adversary]
    For an IMDP $\mathcal{I}=(S,A,\alpha,c,c_h,$ $\underline{P},\overline{P},h)$, an adversary is defined as a map $\xi : \{0,1,\dots,h-1\}\times S\times A\rightarrow \mathsf{P}^{|S|}$. Hence, an adversary $\xi$ is a function that, given the state $s\in S$ and action $a\in A$, at time step $k\in\{0,1,\dots,h-1\}$, selects a feasible transition probability distribution $p\in\mathcal{P}_s^{a}\subseteq\mathsf{P}^{|S|}$. The set of all adversaries is denoted by $\Xi$.
\end{definition}

In the following, we will slightly abuse notation and write $\pi_k(s):=\pi(k,s)$ and $\xi_k^{a}(s):=\xi(k,s,a)$. As with policies, we only consider Markov adversaries, and leave extensions to non-Markovian ones for future work.

Given an IMDP, a policy $\pi$ and an adversary $\xi$, state transitions occur as follows. At time $k$, given the current state $s_k$, an action $a_k$ is randomly selected according to the corresponding probability distribution $\pi_k(s_k)$ defined by policy $\pi$. Then, the adversary $\xi$, given the state $s_k$, chooses a feasible
distribution $p_k^{s_k}:=\xi_k^{s_k}(a_k)\in\mathcal{P}_{s_k}^{a_k}$. The next state of the path $s_{k+1}$ is sampled randomly from $p_k^{s_k}$.

An IMDP $\mathcal{I}$ subject to an adversary $\xi\in\Xi$ and a policy $\pi\in \Pi$ thus simplifies to a time-varying Markov chain (MC), with transition probability matrix 
\begin{equation}
\textstyle P^{\pi,\xi}_k:=\begin{bmatrix}
p_k^{1} & p_k^{2} & \cdots & p_k^{|S|}
\end{bmatrix}, \ \ \ p_{k}^{s}:=\sum_{a\in A}\pi_k^{a}(s)\xi_k^{a}(s), \label{eq:MC_reduction}
\end{equation}
at time $k\in\{0,1,\dots,h-1\}$, where we let $P^{\pi,\xi}:=(P^{\pi,\xi}_0,P^{\pi,\xi}_1,\dots,P^{\pi,\xi}_{h-1})$. Let us use notation $\mathcal{I}^{\pi,\xi}:=(S,\alpha,c,$ $c_h,P^{\pi,\xi},h)$ to refer to the MC that results from the application of policy $\pi$ and adversary $\xi$ to IMDP $\mathcal{I}$, and $X_k \sim \mathcal{I}^{\pi,\xi}$ to refer to a trajectory of the process generated by this MC.

\subsection{Markov process entropy} \label{sec:entropy_of_markov_processes}

In the context of information theory, the concept of \emph{entropy} \cite{Cover1991ElementsOI} describes the degree of uncertainty inherent to the outcome of a random variable. The entropy of a single random variable $X$, which takes values on a finite set $S$, distributed according to $p\in\mathsf{P}^{|S|}$, is often defined as $H(X)=-\sum_{s\in S}p_s\log p_s$, where we use notation $\log:=\log_2$ and we let $x\log x = 0$ for $x=0$ as $\lim_{x\downarrow 0}x\log x=0$. The entropy of a sequence of $h+1$ random variables on $S$ as $X_0,X_1,\dots,X_h$ (possibly for $h\rightarrow \infty$) is described by the joint entropy
\begin{equation}
H(X_0,\dots,X_h)= -\smashoperator{\sum_{s_0s_1\cdots s_h\in S^{h+1}}} p_{s_0s_1\cdots s_h}\log p_{s_0s_1\cdots s_h}, \label{eq:MC_entropy}
\end{equation}
where $p_{s_0s_1\cdots s_h}:=\operatorname{Prob}[X_0=s_0,X_1=s_1,\dots,X_h=s_h]$ denotes the probability measure over the sequences $X_0,X_1,\dots,X_h$.

Due to the Markov property, for sequences of random variables generated by a Markov process $X_k$ over a state space $S$, we might thus alternatively write the $p_{s_0s_1\dots s_h}$ term found in (\ref{eq:MC_entropy}), as 
\begin{equation}
\begin{split}
 p&_{s_0s_1\cdots s_h}= \\
&\operatorname{Prob}[X_0=s_0]\prod_{k=0}^{h-1}\operatorname{Prob}[X_{k+1}=s_{k+1} | X_{k} = s_{k}].\label{eq:MC_entropy_p(s0s1...sh)}
\end{split}
\end{equation}
In the context of IMDPs, these transition probabilities are thus constrained to lie in the interval
\begin{equation}
\operatorname{Prob}[X_{k+1}=q|X_k=s]\in [\underline{P}(s,a,q),\overline{P}(s,a,q)], \label{eq:prob_bounds_IMDP}
\end{equation}
for $k\in\{0,1,\dots,h-1\}$, $s,q\in S$, and action choice $a\in A$.

\subsection{Problem statement}\label{sec:problem_statement}

Let us introduce the shorthand notation $H(\mathcal{I}^{\pi,\xi}):=H(X_0,\dots,X_h \mid X_k\sim \mathcal{I}^{\pi,\xi}, k\in\{0,1,\dots,h\})$, to describe the entropy (\ref{eq:MC_entropy}) of sequence $X_0,\dots,X_h$ generated by IMDP $\mathcal{I}$ subject to policy $\pi$ and adversary $\xi$.

Motivated by real-world scenarios, where predictability of autonomous systems is crucial (e.g., human-robot interaction), we search for policies $\pi\in \Pi$ that, when applied to IMDP $\mathcal{I}$, minimize the cumulative expected cost
\begin{equation}
\textstyle J^{\pi,\xi}=\mathbb{E}[\sum_{k=0}^{h}c(X_k,a_k)], \label{eq:cum_cost}
\end{equation}
where $X_k\sim \mathcal{I}^{\pi,\xi}$, while at the same time keep entropy low. More formally, we are interested in finding the policy $\pi^{*}\in \Pi$ that minimizes the upper bound w.r.t.\ all adversaries $\xi\in \Xi$ on the cost-entropy trade-off of the IMDP
\begin{equation}
\textstyle \overline{J}^{*}(\mathcal{I}):=\min_{\pi\in \Pi}\max_{\xi\in \Xi} J^{\pi,\xi}+\beta H(\mathcal{I}^{\pi,\xi}),\label{eq:main_problem_statement}
\end{equation}
where $\beta\in\mathbb{R}_{\ge 0}$ is a weight factor that tunes the cost vs.\ predictability trade-off. Without loss of generality, it is assumed that $\beta=1$.

In the sequel, we prove that $\overline{J}^{*}$ and an optimal policy can be obtained through value iteration. We additionally show that a deterministic optimizing policy exists. Lastly, we show how the value iteration computations can be solved efficiently through convex optimization.

\section{ROBUST IMDP COST-ENTROPY MINIMIZATION} \label{sec:mainresultI}
In this section we provide a key result (Theorem \ref{thm:minmax_IMDP_recursion}), which shows that $\overline{J}^{*}$ and the corresponding optimal policies can be computed through \emph{value iteration} \cite{Bertsekas1995DynamicPA}. Before presenting this result, we first introduce two lemmas showcasing that both the cumulative cost and the entropy, for an IMDP with given policy $\pi$ and adversary $\xi$, can be computed separately through value iteration.

The first lemma is a celebrated result from standard MDP theory \cite{Bertsekas1995DynamicPA}, here placed in the context of IMDPs. The lemma shows that the expected cumulative cost associated to $\mathcal{I}^{\pi,\xi}$ can be computed via a recursion. The proofs to our results can be found in Section \ref{sec:proofs}, unless otherwise stated.

\begin{lemma}(Recursive Expected Cost Computation)\label{lem:recursive_cost}
    The expected cumulative cost (\ref{eq:cum_cost}) associated with $\mathcal{I}^{\pi,\xi}=(S,\alpha,c,c_h,P^{\pi,\xi},h)$, is given by
    \[
    \textstyle J^{\pi,\xi}= \sum_{s\in S}\operatorname{Prob}[X_0=s]U_0^{\pi,\xi}(s),
    \]
    where $U_0^{\pi,\xi}$ is defined by the recursion
    \begin{equation}
    \textstyle U_k^{\pi,\xi}(s) = \sum_{a\in A}\pi_k^{a}(s)c(s,a) + \sum_{q\in S}p_k^{s}(q)U_{k+1}^{\pi,\xi}(q),\label{eq:cum_cost_recursion}
    \end{equation}
    with initialization $U_h^{\pi,\xi}(s)=c_h(s)$, for $s\in S$, $k\in\{h-1,h-2,\dots,0\}$.
\end{lemma}

\textit{Proof of Lemma \ref{lem:recursive_cost}:}
Follows directly from standard dynamic programming theory \cite{Bertsekas1995DynamicPA}.\qed 

The next lemma shows that entropy can be computed through a similar recursion. This result reflects some aspects of \cite{Chen2014OnTC}, which treats the infinite-horizon IMC entropy maximization. However, here, we present and prove the result for finite-horizon IMDP entropy computation, and later robust minimization using different arguments.

Let function $\Phi : \mathsf{P}^{|S|}\times \mathbb{R}^{|S|}\rightarrow \mathbb{R}$ be defined as 
\begin{equation}
\textstyle \Phi(p,V) := -\sum_{q\in S}p_q\log p_q + \sum_{q\in S}p_q V_q,\label{eq:Phi}
\end{equation}
and $H^{\pi,\xi}(X_{i},\dots,X_{j}):=H^{\pi,\xi}(X_{i},\dots,X_{j}|X_k\sim \mathcal{I}^{\pi,\xi}, k\in\{i,i+1,\dots,j\})$ for some $i,j\in\{0,1,\dots,h-1\}$, $i\le j$.

\begin{lemma}[Recursive Entropy Computation] \label{lem:recursion_of_entropy}   
The entropy of the sequence $X_0,\dots,X_h$ generated according to $\mathcal{I}^{\pi,\xi}=(S,\alpha,c,c_h,P^{\pi,\xi},h)$, is given by 
\begin{equation}
H^{\pi,\xi}(X_0,\dots,X_{h}) = H(X_0) + \sum_{s\in S}\operatorname{Prob}[X_0=s]W_0^{\pi,\xi}(s),\label{eq:recursion_substitution_entropy}
\end{equation}
where $W_0^{\pi,\xi}$ is defined by the recursion
\begin{equation}
W_{k}^{\pi,\xi}(s) =\Phi(p_k^s,W_{k+1}^{\pi,\xi}),\label{eq:recursive_entropy}
\end{equation}
with initialization $W_h^{\pi,\xi}(s)=0$, for $s\in S$, $k\in\{h-1,h-2,\dots,0\}$.  
\end{lemma}

The main result of this section is given next.

\begin{theorem}[Cost-Entropy Trade-Off Minimization]\label{thm:minmax_IMDP_recursion} Given an IMDP $\mathcal{I}:=(S,A,\alpha,c,c_h,\underline{P},\overline{P},h)$, $\overline{J}^{*}(\mathcal{I})$ is given by
\begin{equation}
    \textstyle \overline{J}^{*}(\mathcal{I})=H(X_0)+\sum_{s\in S}\operatorname{Prob}[X_0=s]\overline{V}^{*}_0(s), \label{eq:opt_entropy_bound}
\end{equation}
where $\overline{V}_0^{*}(s)$, $s\in S$, is obtained through the recursion
\begin{equation}
\overline{V}^{*}_{k}(s) = \min_{\pi_k\in \mathsf{P}^{|A|}}\max_{p^{a}\in\mathcal{P}^a_s}\sum_{a\in A}\pi_k^{a}c(s,a)+\Phi(\sum_{a\in A}\pi_k^{a}p^{a},\overline{V}^{*}_{k+1}), \label{eq:minmax_recursion}
\end{equation}
with initialization $\overline{V}^{*}_h(s)=c_h(s)$, for all $s\in S$, $k\in\{h-1,h-2,\dots,0\}$.

Additionally, the optimal policies and adversaries are given by
\begin{align}
    {\xi^{a,*}_k}(s)&\in\smashoperator{{\arg\max}_{p\in\mathcal{P}^a_s}} \ c(s,a)+\Phi(p,\overline{V}^{*}_{k+1}),\label{eq:opt_adversary}  \\
    \pi^*_k(s)&\in\arg\smashoperator{\min_{\pi_k\in \mathsf{P}^{|A|}}} \ \sum_{a\in A}\pi_k^{a}c(s,a)+\Phi(\sum_{a\in A}\pi_k^{a}\xi_k^{a,*}(s), \overline{V}^{*}_{k+1}),\label{eq:opt_policy}
\end{align} 
for all $s\in S$, $a\in A$, $k\in\{0,\dots,h-1\}$.
\end{theorem}

Note that this bound is tight, as the procedure constructs the worst-case adversary (\ref{eq:opt_adversary}), which, under the optimal policy (\ref{eq:opt_policy}) realizes this exact cost-entropy trade-off value.

\section{EFFICIENT VALUE-ITERATION ALGORITHM} \label{sec:mainresultII}
Let the set $\mathsf{P}^{|A|}_{\delta}\subseteq\mathsf{P}^{|A|}$ be the restriction of set $\mathsf{P}^{|A|}$ to the set of indicator vectors $\mathsf{P}_{\delta}^{|A|}:=\mathsf{P}^{|A|}\cap \{0,1\}^{|A|}$. Let us also define $\Pi_{\delta}\subseteq\Pi$ as the set of all policies $\pi_{\delta}$ which deterministically select a single action $a\in A$ at every $k\in\{0,1,\dots,h-1\}$, $s\in S$ as $\pi_{\delta} : \{0,1,\dots,h-1\}\times S \rightarrow \mathsf{P}^{|S|}_{\delta}$.

Intuitively, introducing additional randomness in the effort of entropy reduction is likely counter productive. In fact, we are able to show below that there always exists a \textit{deterministic} policy $\pi_{\delta}\in\Pi_{\delta}$ that realizes the same $\overline{J}^{*}(\mathcal{I})$ as any optimal stochastic policy.

\begin{theorem}[Deterministic Policies Minimize $\overline{J}^{*}(\mathcal{I})$] \label{thm:policy_determinism}
Given an IMDP $\mathcal{I}=(S,A,\alpha,c,c_h,\underline{P},\overline{P},h)$, there exists a deterministic policy $\pi_{\delta}\in\Pi_{\delta}$, such that $\pi_{\delta} \in\arg\min_{\pi\in \Pi} \max_{\xi\in \Xi} J^{\pi\xi}+ H(\mathcal{I}^{\pi\xi})$.

As a consequence, $\overline{V}^{*}_0$ from (\ref{eq:opt_entropy_bound}) can be computed through the recursion
\begin{equation}
\textstyle \overline{V}^{*}_{k}(s) = \min_{a\in A}\max_{p^{a}\in\mathcal{P}_s^{a}}c(s,a)+\Phi(p^{a},\overline{V}^{*}_{k+1}),\label{eq:simpler_recursion}
\end{equation}
with initialization $\overline{V}^{*}_h(s)=c_h(s)$, for all $s\in S$, $k\in\{h-1,h-2,\dots,0\}$, while the corresponding optimal policies and adversaries are found as:
\begin{align*}
    \textstyle {\xi^{a,*}_k}(s)&\textstyle \in \arg\max_{p\in\mathcal{P}^a_s} c(s,a)+\Phi(p,\overline{V}^{*}_{k+1}), \ \ a\in A, \\
    \mu_k^{*}(s)&\in\arg\hspace{0.2mm}\textstyle \min_{a\in A}\hspace{0.45mm} c(s,a)+\Phi(\xi_k^{a,*}(s), \overline{V}^{*}_{k+1}),
\end{align*}
where, through mapping $\mu^*:\{0,1,\dots,h-1\}\times S\rightarrow A$, we find $\pi^*_{\delta,k}(s):=\{p\in\mathsf{P}^{|A|}_{\delta}:p_{\mu_k(s)}=1\}$, for $k\in\{0,1,\dots,h-1\}$, $s\in S$.

Furthermore, the inner max problem in (\ref{eq:simpler_recursion}) is convex, and thus the min-max problems are equivalent to $|A|$ convex programs. 

\end{theorem}


Theorem \ref{thm:policy_determinism} gives rise to Algorithm \ref{alg:mu_algo}, which offers an efficient implementation as a finite number of convex programs, growing linearly with the size of the state and action spaces ($|S|\cdot |A|\cdot h$ to be exact).

\begin{algorithm}
\caption{Efficient Computation of $\mu^{*}$ and $\overline{J}^{*}$} 
\label{alg:mu_algo}
\begin{algorithmic}
    \State Given an IMDP $\mathcal{I}:=(S,A,\alpha,c,c_h,\underline{P},\overline{P},h)$:
    
    \State 1. Set $\overline{V}^{*}_h(s)=c_h(s)$, for all $s\in S$.

    \State 2. For $k\in\{h-1,h-2,\dots,0\}$, via convex optimization, compute for all $s\in S$:
    \begin{align*}
    {\xi^{a,*}_k}(s)&\gets\textstyle \arg\max_{p\in\mathcal{P}^a_s} c(s,a)+\Phi(p,\overline{V}^{*}_{k+1}), \ \ a\in A, \\
    \mu_k^{*}(s)&\gets \arg\hspace{0.2mm}\textstyle \min_{a\in A}\hspace{0.45mm} c(s,a)+\Phi(\xi_k^{a,*}(s), \overline{V}^{*}_{k+1}),\\
    \textstyle \overline{V}^{*}_{k}(s) &\textstyle \gets c(s,\mu_k^{*}(s))+\Phi(\xi_k^{a,*}(s),\overline{V}^{*}_{k+1}). 
    \end{align*}
    \State 3.\hspace{2.6mm} $\overline{J}^{*}(\mathcal{I})\gets\Phi(\alpha,0)+\sum_{s\in S}\alpha(s)\overline{V}^{*}_0(s)$.
\end{algorithmic}
\end{algorithm}

\begin{remark}
In the standard IMDP setting where only a cumulative cost is considered, the inner maximization problem is a linear program \cite{Givan2000BoundedparameterMD,Delimpaltadakis2022IntervalMD}. Due to the additional entropy term, which directly depends on the probability distribution, the inner maximization problem is convex and not linear.
\end{remark}


\section{EXAMPLE APPLICATION} \label{sec:application}
\def\svgwidth{0.4\textwidth} 
\begin{figure}[b]
    \centering
    \input{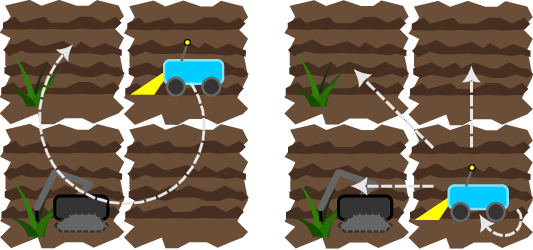}
    \caption{Left: inspection robot A can progress deterministically in a clockwise fashion while exterminator B is not in its way. Right: when B is present in the quadrant ahead of robot A, A makes a highly unpredictable evasive maneuver.}
    \label{fig:illustration}
\end{figure}

In this section, we employ a highly simplified mobile robotics problem in agriculture to demonstrate the efficacy of the tools developed above. After introducing the problem, we use the technique suggested by Algorithm \ref{thm:minmax_IMDP_recursion} to compute the optimal policy and the associated cost-entropy upper-bound to a set of example scenarios.  

Although our approach clearly applies to more general and complex scenarios, let us model an agricultural field as a $2\times 2$ grid, see Fig.\ \ref{fig:illustration}. On this field, a mobile robot (type A) is tasked with monitoring the field by continuously moving over the four quadrants in a clockwise fashion, deterministically moving one grid element at every time-step. 

It is known that the west field quadrants (Q1 and Q2 in Fig.\ \ref{fig:illustration}) are susceptible to weed infections, and the chance of weeds appearing at any time in either of the two western quadrants is found to be in the interval $[0.05,0.5]$. Weed infections are costly as they compete with the crops for nutrients. In order to combat weed infections, a weed exterminator robot (type B) is introduced, also visualised in Fig.\ \ref{fig:illustration}. The type B robot instantly removes weeds in the quadrant on which it is told to act. At every time-step, the quadrant on which robot B acts can be selected $A:=\{1,2,3,4\}$. 

In order to avoid collisions, robot A evades robot B when it acts on the quadrant which robot A intends to cover next. These evasive maneuvers are highly unpredictable and cause robot A to land in any of the four quadrants with a probability between $[0,0.8]$, interrupting its clockwise path. Although the type B robot originally only aims at minimizing weeds, the farmer wishes for it to additionally take into account its effects on the path of robot A, as randomization of the path of robot A causes inconsistencies in the data it collects.

We thus summarize each state as $x:=[l_A \ w_1 \ w_2]^\top$, where $l_A\in\{1,2,3,4\}$ is the location of robot A, $w_1\in\{0,1\}$ is the weed infection status of quadrant 1, and $w_2\in\{0,1\}$ that of quadrant 2. Alternatively, we might label each of the 16 unique sates as $S\in\{1,2,\dots,16\}$. Let each weed infection be associated to a cost of $1$, cumulating each time-step in which it is present, as 
\begin{equation}
c(s,a) = \begin{cases}
    2, &\text{if }s\text{ corresponds to two infected quadrants,} \\
    1, &\text{if }s\text{ corresponds to one infected quadrant,} \\
    0, &\text{otherwise}.
\end{cases}\label{eq:example_stage_cost}
\end{equation}
We further set $h=8$, $c_h(s):=c(s,1)$, $s\in S$, and the IMDP transition probability intervals according to the description above. As randomness in the path of robot A is undesirable, besides minimizing the aforementioned weed-cost (\ref{eq:example_stage_cost}), we additionally minimize its entropy alongside the cost. We achieve this by setting $\beta=1$ in $(\ref{eq:main_problem_statement})$.

\begin{figure}[t]
    \centering
    \includegraphics[width=0.55\linewidth]{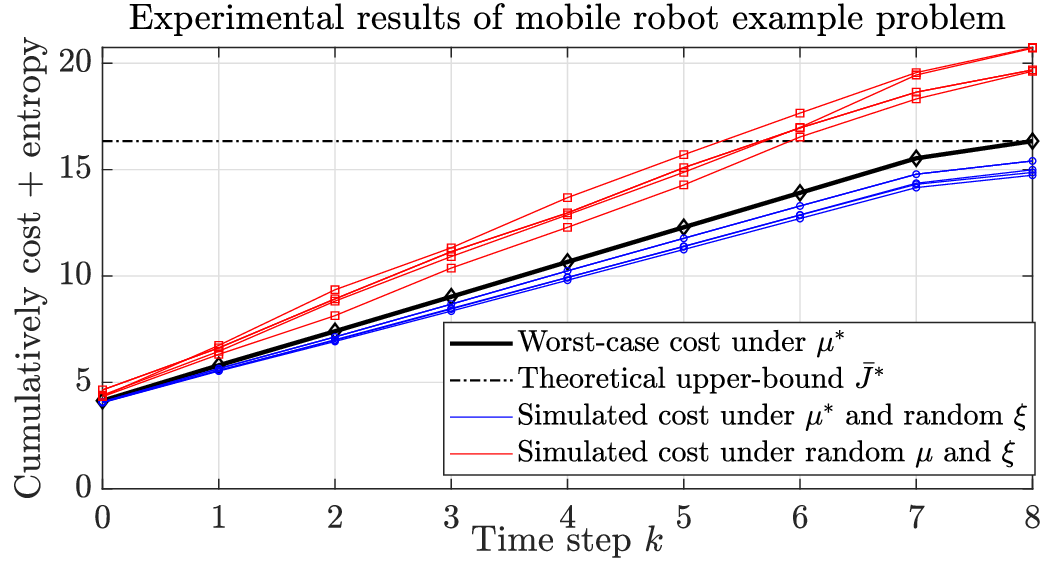}
    \caption{The upper-bound on the linear combination of cumulative cost and entropy under a) the optimal policy and optimal adversary, b) the optimal policy and a random adversary, c) a random policy and random adversary.}
    \label{fig:result}
\end{figure}

Using the optimal policy $\pi^{*}$ and adversary $\xi^{*}$ obtained through the application of Algorithm \ref{alg:mu_algo}, we then simulate the system and visualize the resulting value of cumulative cost and entropy in Fig.\ \ref{fig:result}. There, we compare the value of cumulative cost and entropy associated to the path under optimal policy $\pi^{*}$ and worst-case adversary $\xi^{*}$ in black to (i) the theoretical (tight) upper-bound, which coincides perfectly at $k=h$, (ii) the cumulative cost and entropy associated to a set of paths under the optimal policy $\pi^{*}$ and a random adversary $\xi$ in blue, which, as expected, is lower than the computed upper-bound $\overline{J}^{*}$, and (iii) the cumulative cost and entropy associated to a set of paths under arbitrary policies and adversaries in red, which --- in this specific example --- clearly perform worse than the optimal policy, even when the optimal policy is subjected to $\xi^{*}$.

\begin{figure}[t]
    \centering
    \includegraphics[width=0.48\linewidth]{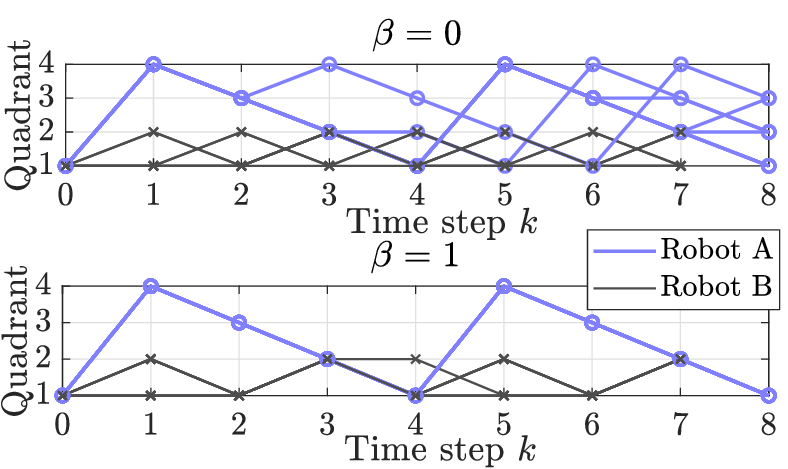}
     \caption{The locations of robots A and B over time in ten simulated trajectories subject to an optimal policy with no entropy regularization (top figure, $\beta=0$), and ten simulated trajectories subject to an optimal policy with entropy regularization (bottom figure, $\beta=1$). We see that regularization of the policy using entropy has the clear effect of improving the predictability of the system.}
    \label{fig:result_trajectories}
\end{figure}

To further demonstrate the effect of entropy regularization on the resulting system behavior, another policy has been computed using Algorithm \ref{alg:mu_algo}, subject to $\beta=0$, i.e., with no entropy regularization. We compare trajectories subject to each of these two optimal policies in Fig.\ \ref{fig:result_trajectories}. There, it becomes clear that the introduction of the entropy regularization term ($\beta\ne 0$) causes a significant increase in the predictability of the system. The policy corresponding to $\beta=1$ yields perfectly predictable robot A behavior, as robot A indeed moves clockwise at every single run of the simulation. In contrast, the policy with no entropy regularization results in the robot A performing many evasive maneuvers and following different trajectories in different runs of the simulation. In fact, these significant gains in predictability come with only a slight loss on optimality w.r.t. the cumulative cost. Specifically, the upper-bound on the expected cumulative cost $J^{\pi^*,\xi}$ associated to the entropy-regularized policy ($\beta=1$) is $7.5619$; only slightly larger than the corresponding bound for the non-regularized policy ($\beta=0$), which is $7.368$.

\section{CONCLUSIONS AND FUTURE WORK} \label{sec:conclusion}
We have shown that robust minimization of the linear combination of entropy and a standard cumulative cost in IMDPs can be solved through value iteration, and that optimal deterministic policies exist. Our value iteration algorithm solves $|S|\times|A|$ convex programs in each time step. 

Possible future research directions include extending the methods described here to cover the infinite-horizon scenario, the maximization of entropy and entropy rate, and investigating questions surrounding its game-theoretic aspects. Furthermore, it would be interesting to address the question of whether, when abstracting stochastic systems through IMDPs, we can get formal guarantees on the predictability of the underlying stochastic dynamical system.

\section{TECHNICAL RESULTS AND PROOFS} \label{sec:proofs}
In this section, we collected the proofs of Theorem \ref{thm:minmax_IMDP_recursion} and Theorem \ref{thm:policy_determinism}, together with all Lemmas used in their construction. We start by presenting all elements that culminate in the proof of Theorem \ref{thm:minmax_IMDP_recursion} in Section \ref{sec:thm1_proofs}. Next, in Section \ref{sec:thm2_proofs}, the same is done with regards to Theorem \ref{thm:policy_determinism}.

\subsection{Results and lemmas regarding the proof of Theorem \ref{thm:minmax_IMDP_recursion}}\label{sec:thm1_proofs}

\textit{Proof of Lemma \ref{lem:recursion_of_entropy}:}
Let us prove, through induction, that 
\begin{equation}
W_0^{\pi\xi}(s) = H^{\pi\xi}(X_1,\dots,X_h|X_0=s), \  s\in S. \label{eq:recursion_proof_goal}
\end{equation}

If for iteration $k+1$, we have that
\begin{equation}
W_{k+1}^{\pi\xi}(s) = H^{\pi\xi}(X_{k+2},\dots,X_h|X_{k+1}=s), \  s\in S, \label{eq:recursion_induction_condition}
\end{equation}
then for iteration $k$, using (\ref{eq:Phi}) and (\ref{eq:recursive_entropy}), we must have that
\begin{equation}
\begin{split}
W_k^{\pi\xi}(s) &= \underbrace{\textstyle -\sum_{q\in S}p_{k}^{sq}\log p_{k}^{sq}}_{H^{\pi\xi}(X_{k+1}|X_{k}=s)}+\underbrace{\textstyle \sum_{q\in S} p_{k}^{sq} W_{k+1}^{\pi\xi}(q),}_{H^{\pi\xi}(X_{k+2},\dots,X_N|X_{k+1})} \\
&=H^{\pi\xi}(X_{k+1},\dots,X_N|X_{k}=s),
\end{split} \label{eq:recursion_induction}
\end{equation}
since $\sum_{s\in S}p_sH(X_{k+1}|X_k=s)=H(X_{k+1}|X_k)$ and $H(X_{k+2}|X_{k+1})+H(X_{k+1}|X_{k})=H(X_{k+2},X_{k+1}|X_{k})$ for Markov processes \cite{Cover1991ElementsOI}.

Secondly, since $W_{h}^{\pi\xi}(s)=0$, $ s\in S$, we have that
\[
\textstyle W_{h-1}^{\pi\xi}(s)=-\sum_{q\in S}p_{h-1}^{sq}\log p_{h-1}^{sq} = H^{\pi\xi}(X_{h}|X_{h-1}=s),
\]
satisfying (\ref{eq:recursion_induction_condition}) for $k=h-2$. 

Furthermore, since from conditional entropy \cite{Cover1991ElementsOI}, we have both that $H^{\pi\xi}(X_0,\dots,X_h)=H(X_0) + H^{\pi\xi}(X_1,\dots,X_h$ $|X_0)$, and $
H^{\pi\xi}(X_1,\dots,X_h|X_0) = \sum_{s\in S}\operatorname{Prob}[X_0=s]H^{\pi\xi}(X_1,\dots,X_h|X_0=s)$, (\ref{eq:recursion_substitution_entropy}) follows by simply substituting these two relations together with (\ref{eq:recursion_proof_goal}) into (\ref{eq:recursion_substitution_entropy}). \qed

\textit{Proof of Theorem \ref{thm:minmax_IMDP_recursion}:} Assume that for some $k\in\{1,2,\dots,h-1\}$, the following holds
\begin{equation}
\begin{split}
&\overline{V}^{*}_{k+1}(s) = \textstyle \min_{\pi\in \Pi}\max_{\xi\in \Xi}U^{\pi\xi}_{k+1}(s)+W^{\pi\xi}_{k+1}(s),\\
&=\textstyle \min_{\pi_{k+1}\cdots\pi_{h-1}}\textstyle \max_{\xi_{k+1}\cdots\xi_{h-1}}U^{\pi\xi}_{k+1}(s)+W^{\pi\xi}_{k+1}(s),\\
&=:[U^{\pi\xi,*}_{k+1}(s)+W^{\pi\xi,*}_{k+1}(s)], \label{eq:induction_step}
\end{split}
\end{equation}
where, with slight abuse of notation, the second equality makes explicit the fact that the values of $U_{k+1}(s)$ and $W_{k+1}(s)$ are independent of $\pi_0\dots\pi_{k}$ and $\xi_0\dots\xi_{k}$. Then, from (\ref{eq:minmax_recursion}), we get
\[
\begin{split}
&\overline{V}^{*}_k(s)=\min_{\pi_k\in \mathsf{P}^{|A|}}\max_{p\in\mathcal{P}_s^{a}}\sum_{a\in A}\pi_k^{a}c(s,a)+\Phi(\sum_{a\in A}\pi_k^{a}p^{a},\overline{V}^{*}_{k+1}), \\
&\phantom{V}=\min_{\pi_k\in \mathsf{P}^{|A|}}\max_{p\in\mathcal{P}_s^{a}}\sum_{a\in A}\pi_k^{a}c(s,a)+ \sum_{q \in S}\sum_{a\in A}\pi_k^{a}p_q^{a}U^{\pi\xi,*}_{k+1}(q) \\
&\phantom{V}\underbrace{-\sum_{q\in S}\sum_{a\in A}\pi_k^{a}p_q^{a}\log \sum_{a\in A}\pi_k^{a}p_q^{a} + \sum_{q \in S}\sum_{a\in A}\pi_k^{a}p_q^aW^{\pi\xi,*}_{k+1}(q)}_{\Phi(\sum_{a\in A}\pi_k^{a}p^{a},W^{\pi\xi,*}_{k+1})}.
\end{split}
\]
Introducing (\ref{eq:cum_cost_recursion}) and (\ref{eq:recursive_entropy}) to the above equation, we get
\[
\begin{split}
\overline{V}_k^{*}(s)&=\hspace{10mm}\smashoperator{\min_{\pi_{k}\pi_{k+1}\cdots\pi_{h-1}\hspace{5mm}}}\hspace{7mm}\smashoperator{\max_{\hspace{2mm}\xi_{k}\cdots\xi_{h-1}}}\hspace{2mm}U^{\pi\xi}_{k}(s)+W^{\pi\xi}_{k}(s), \\
&=\hspace{14mm}\min_{\pi\in \Pi}\max_{\xi\in \Xi}\hspace{5mm}U^{\pi\xi}_{k}(s)+W^{\pi\xi}_{k}(s).
\end{split}
\]

Now, for $k=h$, it trivially holds that $\overline{V}_h^{*}(s)=c_h(s)=\min_{\pi\in \Pi}\max_{\xi\in\Xi}\underbrace{U_h^{\pi\xi}(s)}_{c_h(s)}+\underbrace{W_h^{\pi\xi}(s)}_{0}$. By induction we thus must have that, since (\ref{eq:induction_step}) holds for $k=h$, it also holds for $k\in\{h-1,h-2,\dots,0\}$, proving that $\overline{V}_0^{*}(s)=\min_{\pi\in \Pi}\max_{\xi\in \Xi}U_0^{\pi\xi}(s)+ W_0^{\pi\xi}(s)$.

Substituting this into (\ref{eq:opt_entropy_bound}) yields $\overline{J}^{*}(\mathcal{I})=H(X_0)+$ $\sum_{s\in S}\operatorname{Prob}[X_0=s][\min_{\pi\in \Pi}\max_{\xi\in \Xi}U_0^{\pi\xi}(s)+ W_0^{\pi\xi}(s)]$ $=\min_{\pi\in \Pi}\max_{\xi\in \Xi}\sum_{s\in S}\operatorname{Prob}[X_0=s]U_0^{\pi\xi}(s)+H(X_0)$ $+\sum_{s\in S}\operatorname{Prob}[X_0=s]W_0^{\pi\xi}(s)=\min_{\pi\in \Pi}\max_{\xi\in \Xi}J^{\pi\xi} + H(\mathcal{I}^{\pi\xi}). \hspace{10mm}\text{(Lemma \ref{lem:recursive_cost}, \ref{lem:recursion_of_entropy})}$\qed

\subsection{Results and lemmas regarding the proof of Theorem \ref{thm:policy_determinism}}\label{sec:thm2_proofs}

\begin{lemma}[Concavity of $\Phi(p,V)$]\label{lem:cost_concavity}
The function $\Phi(p,V)$ (\ref{eq:Phi}) is strictly concave w.r.t.\ vector $p\in\mathsf{P}^{|S|}$, meaning that the following inequality holds
\begin{equation}
\Phi(\Sigma {}_{i=1}^{N}\alpha_ip_i,V) \ge \Sigma {}_{i=1}^{N}\alpha_i \Phi(p_i,V), \label{eq:concavity_inequality}
\end{equation}
for any fixed $V\in\mathbb{R}^{|S|}$, $\alpha\in\mathsf{P}^{N}$, and $p_i\in\mathsf{P}^{|S|}$ for $i\in\{1,\dots,N\}$.\qed
\end{lemma}

\textit{Proof of Lemma \ref{lem:cost_concavity}:} Let us rewrite (\ref{eq:Phi}) as $\Phi(p,V)=\sum_{q\in S}\left[-p_q \log p_q + p_q V_q\right]$. We then isolate the summation components for $q\in S$ as $f_q(p_q) := -p_q\log p_q + p_qV_q$, for which we find that $f_q''(p_q) = -(p_q\ln{2})^{-1}<0$ for all non-negative $p_q$, as $\lim_{p_q\rightarrow 0^{+}}-1/(p_q\ln{2})=-\infty$, i.e., $f''(p_q)$ is strictly concave in $p_q\ge 0$. As $\Phi(p,V)$ is thus a sum of strictly concave functions in $p_q\ge 0$ for $q\in S$, $\Phi(p,V)$ itself must be strictly concave in $p\in\mathsf{P}^{|S|}$ (inspired by \cite{Cover1991ElementsOI}, Thm.\ 2.7.1). \qed


\textit{Proof of Theorem \ref{thm:policy_determinism}:} From the fact that $\mathcal{P}_\delta^{|A|}\subset \mathcal{P}^{|A|}$, the following inequality clearly holds
\begin{equation}
\begin{split}
\overline{V}_k^{*}(s)=\min_{\pi_k\in\mathsf{P}^{|A|}}\max_{p^{a}\in\mathcal{P}^{a}}\sum_{a\in A}\pi_k^{a}c(s,a) + \Phi(\sum_{a\in A}\pi_k^{a}p^{a},\overline{V}^*_{k+1})& \\
\le\min_{\pi_k\in \mathsf{P}_{\delta}^{|A|}}\max_{p^{a}\in\mathcal{P}^{a}}\sum_{a\in A}\pi_k^{a}c(s,a) + \Phi(\sum_{a\in A}\pi_k^{a}p^{a},\overline{V}^*_{k+1})& \\
= \min_{a\in A}\max_{p^{a}\in\mathcal{P}^{a}}c(s,a)+\Phi(p_a,\overline{V}^*_{k+1})&,
\end{split} \label{eq:inequality_minimax}
\end{equation}
We show that the opposite inequality holds, thereby confirming (\ref{eq:simpler_recursion}). Note that the following holds for any $\pi_k\in\mathsf{P}^{|A|}$
\begin{equation}
\begin{aligned}
\max_{p^{a}\in\mathcal{P}^{a}}\sum_{a\in A}\pi_k^{a} c(s,a) + &\Phi(\sum_{a\in A}\pi_k^{a} p_a, \overline{V}^*_{k+1}) \\
\ge \max_{p^{a}\in\mathcal{P}^{a}}\sum_{a\in A}\pi_k^{a} c(s,a) + &\sum_{a\in A}\pi_k^{a}\Phi( p_a, \overline{V}^*_{k+1}) \hspace{4.5mm}  \text{(Lemma \ref{lem:cost_concavity})}\\
=\sum_{a\in A}\pi_k^{a} \max_{p^{a}\in\mathcal{P}^{a}}c(s,a) + &\Phi(p_a, \overline{V}^*_{k+1}) \\
\ge \min_{a\in A}\max_{p^{a}\in\mathcal{P}^{a}}c(s,a)+&\Phi(p_a,\overline{V}^*_{k+1}), 
\end{aligned} \label{eq:second_inequality_proof}
\end{equation}
where we used the fact that for any finite sequence of scalars $(\phi_1,\phi_2,\dots,\phi_{|A|})\in\mathbb{R}^{|A|}$, it holds that $\sum_{a\in A}\pi_a \phi_a\ge \min_{a\in A}\{\phi_a\}$, $\forall \pi\in\mathsf{P}^{|A|}$. As (\ref{eq:second_inequality_proof}) holds for any $\pi_k\in\mathsf{P}^{|A|}$, it must also hold for the minimizer of the first line in (\ref{eq:inequality_minimax}).

From this, we conclude that for every iteration of (\ref{eq:minmax_recursion}), minimizing over $a\in A$ will yield the same $\overline{V}^{*}_k(s)$ as the minimization over $\pi_k\in\mathsf{P}^{|A|}$, thus (\ref{eq:main_problem_statement}) minimized over $\mu\in M$, where $M$ is the set of all deterministic policies, instead of $\pi\in\Pi$ will not change the value of $\overline{J}^{*}(\mathcal{I})$ obtained. 

Lastly, from Lemma \ref{lem:cost_concavity} and the fact that sets $\mathcal{P}_s^{a}$ are convex polytopes, we have that the inner-optimization (maximization) in (\ref{eq:simpler_recursion}), for every $a\in A$ is a convex program. \qed



\bibliographystyle{IEEEtran}
\bibliography{mybibliography}

\end{document}